\documentstyle[prl,aps,floats]{revtex}
\begin{document}
\twocolumn[
\title{{\bf Universal Equilibrium Currents in the Quantum Hall Fluid}}
\author{Michael R. Geller and Giovanni Vignale}
\address{Institute for Theoretical Physics, University of California, Santa
Barbara, California 93106}
\address{and Department of Physics, University of Missouri, Columbia, Missouri
65211}
\date{\today}
\widetext\leftskip=0.10753\textwidth \rightskip\leftskip

\begin{abstract}
The equilibrium current distribution in a quantum Hall fluid that
is subjected to a slowly varying confining potential is shown to generally
consist
of strips or channels of current, which alternate in direction, and which have
universal integrated strengths.
A measurement of these currents would yield direct independent measurements of
the proper
quasiparticle and quasihole energies in the fractional quantum Hall states.
\end{abstract}

\bgroup
\draft
\egroup
\maketitle]

Since the discovery of the integral
and fractional quantum Hall effects, a
tremendous experimental and theoretical effort has been made to understand
the nonequilibrium transport current in a two-dimensional (2D) electron gas
subjected to a strong perpendicular magnetic field.
It is now known that the
essential feature leading to the quantization of the Hall conductance is the
existence
of stable incompressible states at certain Landau level
filling factors.
In contrast, relatively
little attention has been given to the {\it equilibrium} current distribution
in a
2D electron gas in the quantum Hall regime, and we shall show here that the
existence of incompressible states leads to some
remarkable properties of the equilibrium current as well.

Central to the progress in understanding the quantum Hall fluid has been the
ability to fabricate semiconductor nanostructures with highly controlled
composition and doping, and the ability to subsequently pattern them, or
provide them
with metal gates, or both. The electron sheet density $\rho$ in a GaAs
heterojunction is typically between $10^{11}$ and $10^{12} \ {\rm cm}^{-2}$.
The quantum Hall regime occurs at low temperatures and at field strengths where
the magnetic length
$\ell \equiv (\hbar c / e B)^{1 \over 2}$ satisfies $\ell \approx \rho^{-{1
\over 2}}$.
At these high field strengths, the magnetic length is often small compared with
the
length scale over which the confining potential---produced by the remote donor
centers
and gates as well as by the electron gas itself---changes by a bulk energy gap.
When this condition is satisfied, the confining potential
is said to be {\it slowly varying}.

In this Letter, we derive a general expression for the low-temperature
equilibrium current distribution in a 2D interacting electron
gas, subjected to a strong perpendicular magnetic field,
and in a slowly varying confining potential. Our expression,
which becomes {\it exact} in the slowly varying limit, has two
components: One is an {\it edge current}, which is proportional to the local
density gradient, and the other is a {\it bulk current}, which is proportional
to the gradient of the self-consistent confining potential. At low
temperatures, the confined
quantum Hall fluid separates into compressible edge channels, where the density
varies,
and incompressible bulk regions, where the density is uniform \cite{edge
channels}.
As we shall show, the screening in the
compressible regions
becomes {\it perfect} at zero temperature, in the sense that the
self-consistent confining potential, which includes exchange and correlation,
becomes constant there.
Hence, the edge current term contributes exclusively to the compressible
regions,
whereas the bulk component contributes only to the
incompressible regions. The resulting current distribution therefore
consists of strips or channels of current, and we shall show that the direction
of
these currents generally displays a striking alternating pattern. Furthermore,
we shall show that the integrated current in each channel is {\it universal},
independent of the width, position, and shape of the channel, as well as the
details of the confining potential.

We begin by calculating the current distribution with current-density
functional
theory, which rigorously accounts for the effects of electron-electron
interaction through
an exchange-correlation scalar potential $V_{\rm xc}$ and vector potential
${\bf A}_{\rm xc}$.
A study of the equilibrium current distribution that neglects exchange
and correlation effects has been reported by us elsewhere \cite{Geller and
Vignale},
and several other authors \cite{currents}
also have contributed to the understanding of equilibrium and
nonequilibrium current distributions.

{\sl I. Equilibrium current distribution}.---Let $H$ be the effective
single-particle
Hamiltonian of current-density functional theory \cite{Vignale and Rasolt},
which contains functionals
of the density $\rho$ and current density ${\bf j}$. We shall use
the self-consistent solutions of the corresponding Kohn-Sham equations,
$H \psi_\alpha = E_\alpha \psi_\alpha ,$
to define an {\it effective} single-particle Green's function for the confined
quantum
Hall fluid,
\begin{equation}
G({\bf r},{\bf r}',s) \equiv \sum_\alpha { \psi_\alpha({\bf r})
\, \psi^*_{\alpha }({\bf r}') \over  s - E_\alpha } ,
\label{greens}
\end{equation}
where $s$ is a complex energy variable
and $\psi$ is a spinor
with components $\psi_\sigma \ (\sigma \! \! \! = \uparrow,\downarrow)$.
Although this effective Green's function
is generally different from the actual single-particle Green's function of the
2D interacting electron gas, it nevertheless yields the {\it exact} equilibrium
number density,
\begin{equation}
\rho({\bf r}) = {\rm Tr} \oint {ds \over 2 \pi i} \ f(s) \ G({\bf r},{\bf
r},s),
\end{equation}
and orbital current density,
\begin{eqnarray}
{\bf j}({\bf r}) = - {e \over m^*} {\rm Tr} \oint {ds \over 2 \pi i}
\ f(s) \ \ \ \ \ \ \ \ \ \ \ \ \ \ \ \ \ \  \nonumber \\
\times \ \lim_{{\bf r}' \rightarrow {\bf r}} \,  {\rm Re}
\, \big(\! -i \hbar \nabla + {e \over c} {\bf A} \big) G({\bf r},{\bf r}',s),
\end{eqnarray}
at fixed chemical potential $\mu$.
Here $e$ is the magnitude of the electron charge, $m^*$ is the effective mass,
and ${\bf A}$ is a vector potential
associated with the uniform external magnetic field ${\bf B}= B {\bf e}_z$.
The contour in the complex energy plane is to be taken in the positive sense
around the poles of $G$, avoiding the poles of the Fermi distribution function
$f(\epsilon) \equiv [e^{(\epsilon - \mu)/k_{\rm \scriptscriptstyle
B}T}+1]^{-1}$.

Next, we let
$V$ be a slowly varying potential from the remote donor centers and gates,
and we write
$H = H^0 + H^1$, where
$H^0 \equiv \Pi^2 /2m^* + {1 \over 2} g \mu_{\scriptscriptstyle \rm B} \sigma_z
B  ,$
${\bf \Pi} \equiv {\bf p} + {e \over c} {\bf A},$ and
\begin{equation}
H^1 \equiv {e \over 2 m^* c} \big( {\bf A}_{\rm xc} \cdot {\bf \Pi} + {\bf \Pi}
\cdot {\bf A}_{\rm xc} \big)
+ V + V_{\scriptscriptstyle \rm H} + V_{\rm xc}.
\label{H1}
\end{equation}
Here $\mu_{\rm \scriptscriptstyle B} \equiv e \hbar / 2 m c$ is the Bohr
magneton,
$V_{\scriptscriptstyle \rm H}$ is the Hartree potential,
$V_{\rm xc}$ is a $2 \! \times \! 2$ diagonal matrix with elements
$ V_{\rm xc}^\sigma \equiv \big( \delta E_{\rm xc}
/ \delta \rho_\sigma  \big)_{\bf v},$ and
\begin{equation}
{\bf A}_{\rm xc} \equiv  - {c \over \rho }
\nabla \times \bigg( { \delta E_{\rm xc} \over
\delta {\bf v} } \bigg)_{\! \rho_\sigma} .
\label{Axc}
\end{equation}
We have omitted a term in $H$ proportional to ${\bf j} \cdot {\bf A}_{\rm xc}$
which is irrelevant in the slowly varying limit.
The exchange-correlation energy $E_{\rm xc}[\rho_\sigma,{\bf v}]$ is a
functional
of the $\rho_\sigma$ and of the gauge-invariant {\it vorticity}
${\bf v} \! \equiv \! \nabla \times \big( {\bf j}_{\rm p} / \rho \big),$
where ${\bf j}_{\rm p}$ is the paramagnetic part of the current density.
Then (\ref{greens}) may be written (again suppressing spin indices) as
\begin{eqnarray}
 G({\bf r},{\bf r}',s) = G^0({\bf r},{\bf r}',s) \ \ \ \ \ \ \ \ \ \ \ \ \ \ \
\ \ \ \ \ \ \ \ \ \ \
\ \ \ \ \ \  \nonumber \\
 + \int d^2r'' \ G^0({\bf r},{\bf r}'',s) \  H^1({\bf r}'')
\ G({\bf r}'',{\bf r}',s) ,
\label{dyson}
\end{eqnarray}
where $G^0$ is the Green's function for the unconfined noninteracting electron
gas.
For large $|{\bf r}-{\bf r}'|,$ the
magnitude of $G^0$ falls off as
$e^{-|{\bf r}-{\bf r}'|^2 \!/4\ell^2} \! \! ,$ except at its poles.

The Dyson
equation (\ref{dyson}) and the short-ranged
nature of $G^0$ may be used to evaluate the effective Green's function
by a {\it gradient expansion} in the self-consistent confining potential.
At each point in the fluid, the confining potential is approximated by a
local potential plus a gradient.
We sum the local confining potential terms to all orders and the gradients to
first order.
The orbital current density is found to be ${\bf j} = {\bf j}_{\rm edge}
+ {\bf j}_{\rm bulk}$, where
\begin{equation}
{\bf j}_{\rm edge}  =  - e \omega_{\rm c} \ell^2 \sum_n (n+{\textstyle{1 \over
2}})
\nabla \rho_n \times {\bf e}_z \ + \ e \omega_{\rm c} \rho
{ {\bf A}_{\rm xc} \over  B },
\label{edge}
\end{equation}
\begin{equation}
{\bf j}_{\rm bulk}  =  - \ {e \over m^* \omega_{\rm c} } \ {\rm Tr}
\ \rho_{\sigma} \ \nabla V_{\rm eff}^{\sigma} \times {\bf e}_z ,
\label{bulk}
\end{equation}
and $\omega_{\rm c} \equiv e B / m^* c$ is the cyclotron frequency. The
electron density is given by
\begin{equation}
\rho = {1 \over 2 \pi \ell^2 } \ \! {\rm Tr}  \ \! \sum_n
f[ (n+{\textstyle{1\ over 2}}
+ {\textstyle {1 \over 2}} \gamma \sigma_z )\hbar \omega_{\rm c} + V_{\rm
eff}({\bf r}) ],
\label{density}
\end{equation}
where $\gamma \equiv g \mu_{\rm \scriptscriptstyle B} B / \hbar \omega_{\rm c}
$ is the dimensionless
spin splitting,
and $\rho_{n}$ is simply the $n$th term in (\ref{density}).
These expressions
differ from those obtained in \cite{Geller and Vignale} by the new edge
current term proportional to ${\bf A}_{\rm xc}$, and by a self-consistent
confining potential,
\begin{equation}
V_{\rm eff} \equiv V + V_{\scriptscriptstyle \rm H} + V_{\rm xc},
\end{equation}
which is modified by exchange and correlation.
Note that no microscopic wave function was needed to obtain (\ref{edge}) and
(\ref{bulk}).
The bulk term (\ref{bulk}) is simply
the transverse Hall current responding to a local
electric field $\nabla V_{\rm eff}/e$.
Discontinuities in $V_{\rm xc}$ will lead to strips of density with fractional
filling factor $\nu \equiv 2 \pi \ell^2 \rho$,
in the same way that the discontinuities in the chemical potential of the
noninteracting
system lead to strips at integral $\nu$.
The total conserved current in a magnetic field
also includes a spin contribution
${\bf j}_{\rm spin} = - {1 \over 2} c g \mu_{\rm \scriptscriptstyle B} \nabla
(\rho_\uparrow
- \rho_\downarrow) \times {\bf e}_z$, which will not be discussed further.

The precise nature of the compressible regions
of the slowly confined quantum Hall fluid as $T \rightarrow 0$ is made
evident by a remarkably simple {\it perfect screening theorem}.
In a slowly varying system, (\ref{edge}) and (\ref{bulk}) may be used to obtain
a total energy functional $E[\rho_\sigma]$ of the density only. Minimizing this
functional
with respect to the $\rho_\sigma$
$(\sigma \! \! = \uparrow,\downarrow)$
under the constraint of fixed total particle number
leads to the conditions
\begin{equation}
\mu_0^\sigma \big( \rho({\bf r}) \big) + V_{\rm eff}^\sigma({\bf r}) = {\rm
constant},
\label{perfect screening}
\end{equation}
where $\mu_0^\sigma (\rho)$ is the spin-$\sigma$ chemical potential of a
noninteracting Hall fluid of uniform
density $\rho$ in the same field ${\bf B}$.
Of course, (\ref{perfect screening}) does not apply
to incompressible regions because $\mu_0$ or $V_{\rm xc}$ is discontinuous
there.
In fact, the discontinuity in $\mu_0 + V_{\rm eff}$ at an incompressible strip
is
precisely equal to the electron chemical potential gap $\Delta \mu$ there.
Because $\mu_0^\sigma(\rho)$ becomes piecewise constant as $T \rightarrow 0$,
{\it the self-consistent potentials} $V_{\rm eff}^\sigma({\bf r})$
{\it in each compressible region must become uniform in this limit}.
Therefore, there exists a {\it complementarity} in the low-temperature phase of
the slowly confined quantum Hall fluid: In the incompressible regions,
$\rho$ is uniform and $V_{\rm eff}$ varies, whereas in the compressible
regions,
$\rho$ varies and $V_{\rm eff}$ is uniform.

Because $\rho$ and ${\bf v}$ are slowly varying, (\ref{Axc}) may be written as
\begin{equation}
{\bf A}_{\rm xc} = \bigg( {m^* c^2 \over e^2 }\bigg) \ { 1 \over \rho}
\ \nabla \times {\bf M}_{\rm xc},
\end{equation}
where ${\bf M}_{\rm xc}$ is the exchange-correlation contribution
to the orbital magnetization of a uniform 2D electron gas.
The second edge
current term in (\ref{edge}) may
therefore be rewritten as $c \nabla \times {\bf M}_{\rm xc}$.

The largest contributions to the current density come from the first term
in (\ref{edge}) and from (\ref{bulk}), and these have opposite signs because
$\nabla \rho$ and $\nabla V_{\rm eff}$ are antiparallel.
Furthermore, the perfect screening in the edge regions makes the bulk
current vanish there,
and the incompressibility of the bulk regions causes the density
gradient and also ${\bf A}_{\rm xc}$ to vanish in those regions.
{\it Therefore, the current distribution generally consists of a series of
strips or
channels of distributed current, which follow the equipotentials of
the self-consistent confining potential, and which alternate in direction.}
We shall show that the origin of this striking alternating pattern is the
oscillations
in the low-temperature magnetization of a 2D electron gas,
in a fixed magnetic field, as a function of density.
These oscillations are of course caused by the same
competition between the energy of a Landau level and its degeneracy that
leads to the deHaas--van Alphen effect.

{\sl II. Integrated equilibrium currents}.---The integrated currents
follow straightforwardly from
(\ref{edge}) and (\ref{bulk}).
For example, the magnitude of the integrated current at the edge of the filled
Landau level $n$,
assuming there is no bulk current present, is
\begin{equation}
I_{\rm edge} = \ (2n + 1) {e \omega_{\rm c} \over 4 \pi}
\ + \ c \, \Delta  M_{\rm xc} ,
\label{Iedge}
\end{equation}
where $\Delta M_{\rm xc}$ is the change in the $z$ component of ${\bf M}_{\rm
xc}$
across the edge channel. Similarly, the integrated current in a bulk region of
filling
factor $\nu \equiv 2 \pi \ell^2 \rho $ has magnitude
\begin{equation}
I_{\rm bulk} = \nu {e \over h} \, \Delta \mu = c \Delta M ,
\label{Ibulk}
\end{equation}
where $\Delta \mu$ is the energy gap and $\Delta M$ the discontinuity
in the magnetization there.
In (\ref{Ibulk}), we have used the fact that, according to the
Maxwell relation
$\big(\partial M / \partial \rho \big)_{\! B}
= - \big( \partial \mu / \partial B \big)_{\! \rho}$,
the discontinuities in $\mu$ and $M$ are generally related by
$\Delta M / \mu_{\rm \scriptscriptstyle B}^* \rho = 2 \Delta \mu / \hbar
\omega_{\rm c}$.

The validity of (\ref{Iedge}) {\it appears} to require that the
channel not have any bulk current
present from possible
incompressible strips at fractional filling factors.
However, we shall prove now
that (\ref{Iedge}) correctly gives the integrated current at the edge of a
filled Landau
level, {\it including} the bulk currents present from strips at fractional
$\nu$.

The equilibrium density is stationary, so $\nabla \cdot {\bf j} = 0$, and
we may write the orbital current as
\begin{equation}
{\bf j} = c \nabla \times M {\bf e}_z ,
\label{curl}
\end{equation}
where $M$ is the $z$ component
of the local thermodynamic magnetization.
Therefore, between any two regions in the
2D electron gas having slow density variation, the integrated equilibrium
current is simply
\begin{equation}
I =  - c \ \Delta M .
\label{magnetization}
\end{equation}
This result is valid
regardless of whether there are additional
incompressible channels present, and
regardless of whether the system is slowly varying
in the intermediate region.

Let $M = M_0 + M_{\rm xc}$,
where $M_0$ is the kinetic contribution.
The ground-state kinetic energy per unit area, ignoring spin, is
${\cal E}_0 = f_0(\nu) \ (\hbar \omega_{\rm c} / 2 \pi \ell^2)$,
where $f_0(\nu) \equiv {1 \over 2} [\nu]^2 + ([\nu]+ {1 \over 2})(\nu -
[\nu])$,
and where $[x]$ denotes the integer part of $x$.
The kinetic chemical potential
$\mu_0 \equiv (\partial {\cal E}_0 / \partial \rho)_{\scriptscriptstyle B}$
is discontinuous at all integral fillings factors by an amount $\hbar
\omega_{\rm c}$.
At $T = 0$,
\begin{equation}
M_0 = \bigg( [\nu] - (2 [\nu] + 1) (\nu - [\nu] ) \bigg)
{\mu^*_{\scriptscriptstyle \rm B} \over 2 \pi \ell^2 },
\label{M0}
\end{equation}
where $\mu^*_{\scriptscriptstyle \rm B} \equiv (m/m^*) \mu_{\rm
\scriptscriptstyle B}$.
Note the discontinuity in
$M_0 / \mu^*_{\scriptscriptstyle \rm B} \rho$ at integral filling factors,
which is equal to
twice the discontinuity in $\mu_0 / \hbar \omega_{\rm c}$ there.
The change in $M_0$ across the
edge of a filled Landau level leads to the first term in (\ref{Iedge}),
and the change in $M_{\rm xc}$ leads to the second term.
Therefore, (\ref{Iedge}) is correct regardless
of whether there are fractional incompressible strips present in the edge
region.
We shall return to this point below.

The interaction energy per area of a quantum Hall
fluid generally depends on $\rho$ and $B$ separately, but with the
assumption of negligible Landau level mixing by the
interactions, we may write
${\cal E}_{\rm xc} = f_{\rm xc}(\nu) \ (e^2 / 2 \pi \kappa \ell^3)$,
where $ \kappa $ is the bulk dielectric constant.
In terms of $f_{\rm xc}$,
\begin{equation}
M_{\rm xc} = \alpha \bigg( -3 f_{\rm xc} + 2 \nu {f'}_{\rm \! \! xc} \bigg)
 {\mu_{\rm \scriptscriptstyle B}^* \over  2 \pi \ell^2},
\end{equation}
where $\alpha \equiv (e^2/\kappa \ell \hbar \omega_{\rm c})$ is the
dimensionless
Coulomb interaction strength.

We now calculate the integrated
current at the edge of the lowest spin-polarized Landau
level $(n \sigma \! = \! 0 \! \downarrow)$, between $\nu = 1^-$ and $\nu = 0$,
to order $\alpha$.
We shall consider for convenience a long Hall bar oriented in the $y$
direction, with a confining potential $V(x)$ that varies in the $x$ direction
only (except near the two ends of the Hall bar, which we avoid). The current
is then directed along the Hall bar in the $y$ direction.
Near $\nu = 0$, the ground-state
energy is expected to be close to that of a Wigner crystal, so
$f_{\rm xc} \propto - \nu^{3 \over 2}$ there.
Particle-hole symmetry implies
$f_{\rm xc}(1-\nu) = f_{\rm xc}(\nu) + (1 - 2 \nu) \, f_{\rm xc}(1)$,
where $f_{\rm xc}(1) = - (\pi/8)^{1 \over 2}.$
Hence, we find $M_{\rm xc}(1^-) = - \alpha (\pi/8)^{1 \over 2}
(\mu_{\scriptscriptstyle \rm B}^* / 2 \pi \ell^2),$
and therefore
\begin{equation}
I_{0 \downarrow} = - \bigg( 1 + \alpha \sqrt{{\pi \over 8}} \ \bigg) {e
\omega_{\rm c} \over 4 \pi} .
\label{filled edge current}
\end{equation}
Expressions for the integrated currents at the edge of higher filled
Landau levels, obtained by a similar analysis, are presented in Table \ref{edge
table}.

\begin{table}
\begin{tabular}
{lcccc}
$n  \!:$ \ \ \ \ \ \ $0$  & $1$ & $2$ & $3$ \\ [.05cm]\hline
\rule[-0.3cm]
{-0.10cm}{0.8cm}
$\ \ 1+\alpha {1\over 2}({\pi \over 2})^{1 \over 2}$
& $3+\alpha {11\over 8}({\pi \over 2})^{1 \over 2}$
& $5+\alpha {265\over 128}({\pi \over 2})^{1 \over 2}$
& $7+\alpha {1371\over 512}({\pi \over 2})^{1 \over 2}$ \\ [0.3cm]
\end{tabular}
\caption{Magnitude of the integrated orbital current at the edge of the filled
Landau level $n$,
apart from a spin-degeneracy factor of 2,
for the case of negligible spin splitting $(\gamma < < 1)$ .
Currents are in  units of $e \omega_{\rm c}/ 4 \pi$.}
\label{edge table}
\end{table}

Next, we shall assume that there is an
incompressible strip at an odd-denominator filling factor
$\nu_{\scriptscriptstyle 0} = {1 \over q}$ present in this edge channel.
At filling factors very close to $\nu_{\scriptscriptstyle 0}$, the ground
state is expected to be a Laughlin
state plus a Wigner crystal of fractionally charged quasiparticles
or quasiholes.
Let $\epsilon_{\scriptscriptstyle L}$ be the interaction energy per electron
in the Laughlin state at $\nu_{\scriptscriptstyle 0}$, in units of $e^2/ \kappa
\ell$,
and let $\epsilon_{\rm qp}$ and $\epsilon_{\rm qh}$ be the associated
gross quasiparticle and quasihole energies in the same
units. Then close to $\nu_{\scriptscriptstyle 0}$,
\begin{equation}
 f_{\rm xc}(\nu) = \bigg\lbrace  \matrix{
\epsilon_{\scriptscriptstyle \rm L} / q + q \epsilon_{\rm  qh}
(\nu_{\scriptscriptstyle 0}-\nu ) + \cdots
\ \ {\rm for} \ \nu \leq \nu_{\scriptscriptstyle 0} \cr
\epsilon_{\scriptscriptstyle \rm L} / q + q \epsilon_{\rm qp} (\nu -
\nu_{\scriptscriptstyle 0}) + \cdots
\ \ {\rm for} \ \nu \geq \nu_{\scriptscriptstyle 0} } .
\label{exchange-correlation energy}
\end{equation}
For example, Morf and Halperin \cite{Morf and Halperin} have evaluated
$\epsilon_{\scriptscriptstyle \rm L}$,  $\epsilon_{\rm qp}$,
and $\epsilon_{\rm qh}$ using trial wave functions
at $\nu_{\scriptscriptstyle 0}
\! = \! {1 \over 3}$;
they find
$\epsilon_{\scriptscriptstyle \rm L} \! = \! - 0.410$,
$\epsilon_{\rm qp} \! = \! -0.132 $,
and $\epsilon_{\rm qh} \! = \! 0.231$.
According to (\ref{exchange-correlation energy}),
$\mu_{\rm xc} \equiv (\partial {\cal E}_{\rm xc}
/ \partial \rho)_{\scriptscriptstyle B}$ is discontinuous by an amount $\Delta
\mu = q E_{\rm gap}$,
where
$ E_{\rm gap} \equiv \epsilon_{\rm qp}
+ \epsilon_{\rm qh} $
is the energy required to create a single well-separated
quasiparticle-quasihole
pair.
The interaction contribution to the orbital magnetization near
$\nu_{\scriptscriptstyle 0}$
is therefore
\begin{equation}
M_{\rm xc} = \bigg\lbrace  \matrix{
- 2 \alpha {\tilde \epsilon}_{\rm qh}
\big( \mu_{\scriptscriptstyle \rm B}^* / 2 \pi \ell^2 \big)
\ \ \ \ {\rm for} \ \nu = \nu_{\scriptscriptstyle 0}^- \cr
\ \ 2 \alpha {\tilde \epsilon}_{\rm qp}
\big( \mu_{\scriptscriptstyle \rm B}^* / 2 \pi \ell^2 \big)
\ \ \ \ {\rm for} \ \nu = \nu_{\scriptscriptstyle 0}^+ } \ ,
\end{equation}
where
${\tilde \epsilon}_{\rm qh} = \epsilon_{\rm qh} + 3
\epsilon_{\scriptscriptstyle \rm L}/2 q$
and
${\tilde \epsilon}_{\rm qp} = \epsilon_{\rm qp} - 3
\epsilon_{\scriptscriptstyle \rm L}/2 q$
are the {\it proper} quasihole and quasiparticle energies defined in \cite{Morf
and Halperin}.
The discontinuity in
$ M_{\rm xc} / \mu_{\rm \scriptscriptstyle B}^* \rho$ at
$\nu_{\scriptscriptstyle 0}$ is equal to twice
the discontinuity in $\mu_{\rm xc} / \hbar \omega_{\rm c}$, as expected.

Let $I_1$ be the integrated current in the edge channel between $\nu = 1^-$
and $\nu = \nu_{\scriptscriptstyle 0}^+$, $I_2$ be the current in the
incompressible strip at
$\nu = \nu_{\scriptscriptstyle 0}$, and $I_3$ be the current in the edge
channel between
$\nu = \nu_{\scriptscriptstyle 0}^-$ and $\nu = 0$. According to (\ref{edge})
or (\ref{magnetization}),
\begin{equation}
I_1 = - \bigg[ 1 - {1 \over q} + \alpha \bigg( \sqrt{\pi \over 8}
+ 2  {\tilde \epsilon}_{\rm qp} \bigg) \bigg] {e \omega_{\rm c} \over 4 \pi}.
\label{I1}
\end{equation}
Similarly, $I_2$ is given by $2 \alpha E_{\rm gap} (e \omega_{\rm c} / 4 \pi).$
Restoring units to $E_{\rm gap}$ leads to
\begin{equation}
I_2 = {e \over h} E_{\rm gap}  = \nu {e \over h} \ \! \Delta \mu ,
\label{I2}
\end{equation}
as in (\ref{Ibulk}).
{\it The magnitude of the integrated equilibrium current in any
incompressible strip is }
$\nu (e/h) \, \Delta \mu$,
{\it where }
$\Delta \mu$
{\it is the electron chemical potential gap in the uniform quantum Hall fluid
at filling factor} $\nu$.
Finally,
\begin{equation}
I_3 = - \bigg[ {1 \over q}
+ 2  \alpha {\tilde \epsilon}_{\rm qh}  \bigg] {e \omega_{\rm c} \over 4 \pi}.
\label{I3}
\end{equation}
Note the alternating signs of the integrated currents, and that
their sum agrees with (\ref{filled edge current}), even though there is now an
incompressible
strip at $\nu_{\scriptscriptstyle 0}$.
Furthermore, a measurement of the currents (\ref{I1}-\ref{I3}) would provide
direct {\it independent} measurements of the fundamental quantities
${\tilde \epsilon}_{\rm qp}$, ${\tilde \epsilon}_{\rm qh}$, and, of course,
$E_{\rm gap}$.

In conclusion, we have derived an expression for the low-temperature
equilibrium
current distribution in a confined quantum Hall fluid. The current distribution
has two components, (\ref{edge}) and ({\ref{bulk}), which contribute
exclusively
to the compressible and incompressible regions respectively, and have opposite
signs. The current distribution therefore consists of strips or channels of
current,
which alternate in direction. The integrated currents in the channels are also
shown
to be universal, and it is noted that their measurement would yield direct
independent measurements
of the proper quasiparticle and quasihole energies in the fractional quantum
Hall states.
Several experimental groups are exploring the possibility of directly imaging
the current
distribution in a 2D electron gas, and preliminary results have already been
reported \cite{Kent etal}.

This work was supported by the NSF through Grants
DMR-9100988 and DMR-9403908. We acknowledge the hospitality of the
Institute for Theoretical Physics,
Santa Barbara, where part of this work was completed under
NSF Grant PHY89-04035.

\end{document}